\documentclass[twocolumn,aps,prb,superscriptaddress,showpacs]{revtex4}
\usepackage{bm}
\usepackage{amsmath}
\usepackage{amssymb}
\usepackage{graphicx}
\usepackage{epsfig}

\begin{document}
\title{Spin-thermo-electronic oscillator based on inverse giant magnetoresistance}
\author{A. M. Kadigrobov}
\affiliation{Department of Physics, University of Gothenburg, SE-412
96 G{\" o}teborg, Sweden} \affiliation{Theoretische Physik III,
Ruhr-Universit\"{a}t Bochum, D-44801 Bochum, Germany}
\author{S. Andersson}\affiliation{Nanostructure Physics, Royal Institute
of Technology, SE-106 91 Stockholm, Sweden}
\author{Hee Chul Park}
\affiliation{Department of Physics, University of Gothenburg, SE-412
96 G{\" o}teborg, Sweden} \affiliation{Department of Physics,
Chungnam National University, Daejeon 305-764, Republic of Korea}
\author{D.~Radi\'{c}}
\affiliation{Department of Physics, University of Gothenburg, SE-412
96 G{\" o}teborg, Sweden} \affiliation{Department of Physics,
Faculty of Science, University of Zagreb, 1001 Zagreb, Croatia}
\author{R. I. Shekhter}
\affiliation{Department of Physics, University of Gothenburg, SE-412
96 G{\" o}teborg, Sweden}
\author{M. Jonson}
\affiliation{Department of Physics, University of Gothenburg, SE-412
96 G{\" o}teborg, Sweden} \affiliation{School of Engineering and
Physical Sciences, Heriot-Watt University, Edinburgh EH14 4AS,
Scotland, UK}\affiliation{ Division of Quantum Phases
and Devices, School of Physics, Konkuk University, Seoul 143-701, Republic of Korea}
\author{V. Korenivski}\affiliation{Nanostructure Physics, Royal Institute
of Technology, SE-106 91 Stockholm, Sweden}

\begin{abstract}
A spin-thermo-electronic valve with the free layer of exchange-spring type and inverse magnetoresistance is investigated. The structure has S-shaped current-voltage characteristics and can exhibit spontaneous oscillations when integrated with a conventional capacitor within a resonator circuit. The frequency of the oscillations can be controlled from essentially dc to the GHz range by the circuit capacitance.
\end{abstract}

\maketitle

\section{Introduction \label{Int}}
Electrons in ferromagnetic conductors  are differentiated with respect to their spin due to the exchange interaction which leads to a splitting of the corresponding energy bands and to a non-vanishing polarization of the conduction electrons. This spin polarization gives rise to a number of  "spintronic" effects in magnetic nanostructures, such as Giant Magnetoresistance (GMR) \cite{Baibich,Binasch} and Spin-Transfer-Torque \cite{Slonczewski,Berger}.

A conventional GMR spin-valve has low resistance in the parallel configuration  of the two ferromagnetic layers and high resistance in the antiparallel configuration. The GMR is said to be inverse when the low resistance state corresponds to the antiparallel orientation of the ferromagnets \cite{Vouille}. Inverse GMR is observed in, for example, spin-valves incorporating minority carrier ferromagnetic layers \cite{George,Renard,Hsu,Voulle2}. Another possible implementation of the inverse GMR spin-valve, demonstrated in this paper, relies on an antiparallel exchange-pinning of two ferromagnetic layers of different coercivity to two antiferromagnetic layers.

We have previously analyzed a novel spin-thermo-electronic (STE) oscillator \cite{JAP} based on a GMR exchange-spring multilayer \cite{Davies} having a N-shaped current-voltage characteristic (IVC), connected in series to an inductor (L) which performs the function of current limiting in the circuit and thereby determines its resonance frequency. A tunable inductor, such as the one having a magnetic core \cite{Vlad1,Vlad2}, can be used to achieve a tunable resonator. However, in applications where the device footprint is tightly budgeted, it may be advantageous to employ a mirror-circuit configuration. Namely, an inverse GMR spin-valve with an S-shaped IVC connected to a capacitor (C) in parallel. This paper provides a detailed analysis of this new structure as well as a comparison of the two STE oscillator designs. We show that the STE-C design proposed herein offers an attractive, compact alternative to the STE-L oscillator discussed previously \cite{JAP}.

%In Section \ref{s-shapedIVC}  we formulate and investigate the transport problem in which  electrical and thermal fluxes through a  magnetic exchange-spring structure with an inverse GMR are interdependent because of the temperature dependence of the relative orientation of the magnetization of the layers \cite{JAP}.

%We show that  the current-voltage characteristic of the stack is  S-shaped and describe in details the way of electrical manipulation of the magnetization direction in the stack.

%In Section \ref{selfoscillations} we study development of the instability of the time-independent state which arises if the stack is incorporated in an electrical circuit in parallel with a capacitor under dc bias current (see Fig.\ref{circuit}) and show that in such a system, nonlinear temporal oscillations of the voltage drop, current, temperature, and magnetization direction in the nanopillar spontaneously arise.

%\section{Thermal-electric manipulation of the magnetization direction \label{manipulation}}
\section{S-shaped current-voltage characteristic under Joule heating: control of the magnetization direction by dc bias current \label{s-shapedIVC}}

\begin{figure}
  %%%%%%%%%%%%%%%%%%%%%%%%%%%%%%%% \centerline{\psfig{figure=zlaser2.eps,width=8cm}}
 %\centerline{\includegraphics[width=0.7\columnwidth]{Stackmagnetization1.eps}}
 \epsfig{file=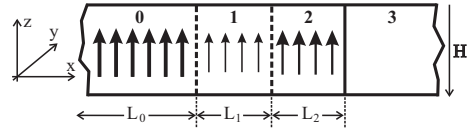, width=0.7\columnwidth}
  %\vspace{1cm}
  \caption{ Orientation of the magnetic moments in
  a  stack of three ferromagnetic layers
(0, 1, 2) in contact with a nonmagnetic layer 3; the magnetic moments in layers 0, 1, and 2 are coupled by the exchange interaction thus building an exchange spring tri-layer; $H$ is an external magnetic field directed antiparallel to the magnetization in layer 0.}
 \label{noflip}
 \end{figure}

We consider a system of three ferromagnetic layers in which two strongly ferromagnetic layers 0 and 2 are exchange coupled through a weakly ferromagnetic spacer (layer 1) while layer 3  is nonmagnetic  as illustrated in Fig.\ref{noflip}. We assume that the Curie temperature $T_c^{(1)}$ of layer 1
is lower than the Curie temperatures $T_c^{(0,2)}$ of
layers 0, 2; we also assume the magnetization
direction of layer 0 to be fixed;
layer 2 is subject to a magnetic field $H$ directed
opposite to the magnetization of layer 0, which can be an
external field, the fringing field from layer 0, or a combination of
the two. We require this magneto-static field  to be weak enough so
that at low temperatures $T$ the magnetization of layer 2 is kept
parallel to the magnetization of layer 0 due to the exchange
interaction between them via layer 1. In the absence of the external
field and if the temperature is above the Curie point of the spacer, $T> T_c^{(1)}$, this tri-layer is similar to the spin-flop `free layer'
widely used in memory device applications\cite{Worledge}.

%%%%%%%%%%%%%%%%%%%%%%%%%%%%%%%%%%%%%%%%%%%%%%%
  %%%%%%%%%%%%%%%%%%%%%%%%%%%%%%%%%%%%%%%%%%%%%%%

As it was shown in paper \cite{JAP}  parallel orientations of the magnetization in layers 0, 1 and 2 becomes unstable if the temperature exceeds some critical temperature $T_c^{\rm(or)} < T_c^{(1)}$.
The magnetization direction in layer 2 tilts  with an increase of the stack temperature $T$ in the temperature range $T_c^{\rm(or)} \leq T \leq T_c^{(1)}$. The dependence of the  equilibrium tilt angle $\Theta$ between the magnetization directions of layers 0 and 2 on $T$ and the magnetic filed $H$ is determined by the equation \cite{JAP}
\begin{eqnarray}\label{theta2}
\Theta=D(H,T)\sin{\Theta}, \hspace{1.0 cm} T < T_c^{(1)} \nonumber \\
\Theta=\pm \pi, \hspace{1.0 cm} T \geq T_c^{(1)}
\end{eqnarray}
where
\begin{equation}
D(H,T)=\frac{L_1 L_2H M_2(T)}{4 \alpha_1 M_1^2(T)}.
 \label{D}
\end{equation}
Here $L_1, \; M_1 $ and $L_2, \; M_2$ are the widths  and the magnetic moments of layers 1 and 2, respectively; $\alpha_1 \sim J_1/ a M_1^2(T=0)$ is the exchange constant, $J_1$ is the exchange energy in layer 1  and $a$ is the lattice spacing. The parameter $D(H,T)$ is the ratio between the
magnetic energy and the energy of the stack volume for the
inhomogeneous distribution of the magnetization.
At low temperatures the exchange energy prevails, the parameter
$D(H,T) <1$ and Eq.~(\ref{theta2}) has only one root,
$\Theta=0$, thus a parallel orientation of  magnetic moments
in layers 0, 1 and 2 of the stack is thermodynamically stable. However,
at temperature $T_c^{\rm(or)} < T_c^{(1)}$, for which
$$D(T_c^{\rm(or)},H)=1,$$
two new solutions $\Theta =\pm |\theta_{\rm
min}| \neq 0$ appear. The parallel magnetization corresponding to
$\Theta=0$ is now unstable, and the  direction of the
magnetization in region 2 tilts   with an increase of temperature inside the interval $T_c^{\rm(or)} \leq T \leq  T_c^{(1)}$.
  According to Eq.(\ref{D}) the critical temperature $T_c^{\rm(or)}$ of this orientational phase transition \cite{Giovanni} is  equal to
\begin{equation}
T_c^{\rm(or)}= T_c^{(1)}\left(1- \frac{\delta T}{T_c^{(1)}} \right), \hspace{0.2cm}
 \frac{\delta T}{T_c^{(1)}}=\frac{L_1 L_2H M_2}{4 \alpha_1 M_1^2(0)}\equiv D_0 \label{Torient}\,.
\end{equation}

If the stack  is Joule heated  by  current $I$ its
temperature $T(V)$   is determined by  the heat-balance condition
\begin{equation}
IV=Q(T), \hspace{0.2cm}I  =V/R(\Theta),
 \label{heat}
\end{equation}
and Eq.~(\ref{theta2}), which determines  the temperature dependence
of $\Theta[T(V)]$. Here  $V$ is the voltage drop across the stack,  $Q(T)$ is the heat flux from the stack and
$R(\Theta)$ is the total stack magnetoresistance.

Equations~(\ref{heat}) and (\ref{theta2}) define the
IVC of the stack
\begin{equation}
I_0(V)=\frac{V}{R[\Theta(V)]},
\label{IVC0}
\end{equation}
where $\Theta(V)\equiv \Theta[T(V)]$.

The differential conductance of the stack \cite{JAP} is
\begin{equation}
\frac{dI_0}{dV}=R \left(\Theta \right) \frac{[R^{-1}\left(\Theta) (1-{\bar
D}\sin{\Theta}/\Theta\right)]'}{[R(\Theta) (1-{\bar
D}\sin{\Theta}/\Theta)]'}\Bigr|_{\Theta=\Theta(V)},
 \label{diffG}
\end{equation}
where
%$${\bar D}=D_0 \frac{T}{Q}\frac{dQ}{dT}\bigr|_{T=T_c^{(1)}}\approx D_0, $$
$[\ldots]'$ means the derivative of the bracketed quantity with
respect to $\Theta$, and $${\bar
D}=\frac{T}{Q}\frac{d Q}{d T} D_0\Bigl|_{T=T_c^{(1)}}.$$

\begin{figure}
%%%%%%%%%%%%%%%%%%%%%%%%%%%%%%%% \centerline{\psfig{figure=zlaser2.eps,width=8cm}}
 %\centerline{\includegraphics[width=0.65\columnwidth]{I-Vst.eps}}
%\vspace{0 mm}
\epsfig{file=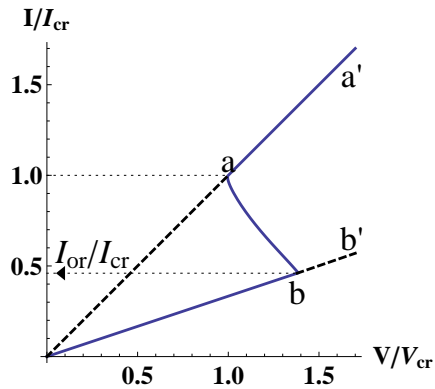, width=0.65\columnwidth}
\caption{ Current-voltage characteristics (IVC) of the
magnetic stack from Fig.~\ref{noflip} calculated for
$R(\Theta)=R_+(1+r \cos\Theta)$, $R(\pi)/R(0)=0.3$, ${\bar
D}=0.36$;
$I_{cr}=\sqrt{Q(T_c^{(1)})/R(\pi)}$ and $I_{or}=\sqrt{Q(T_c^{(or)})/R(0)}$. Branches $0-b$ and $a-a'$  of the IVC
correspond to parallel and antiparallel orientations of the
magnetization of layers 0 and 2, respectively (parts $a-a'$ and $0-b$ are
unstable); branch $a-b$ corresponds to the tilt of the magnetization of layer 2 with respect to that of layer 0 (that is $\Theta \neq 0$).}
 \label{cvc}
\end{figure}

As follows from Eq.(\ref{diffG}) the current-voltage characteristic $I_0(V)$ may be N- or S-shaped depending on whether  the magnetoresistance of the stack  is normal or inverse. In the case that the stack has a normal magnetoresistnace (that is $d R/ d \Theta > 0$) the IVC is N-shaped so nonlinear current and magnetization-direction oscillations may spontaneously arise if the stack is incorporated in a voltage biased electrical circuit in series with an inductor \cite{JAP}. In this paper we consider the situation in which the stack has an inverse GMR \cite{inverse},
i.e. $d R/d\Theta < 0$. As one can see from Eq.(\ref{diffG}), if $d R/d\Theta < 0$, the numerator of the differential conductance is always positive while the denominator can be negative and hence the IVC of the stack is S-shaped as illustrated in Fig.~\ref{cvc}.

For the magnetoresistance of the stack in the form
\begin{equation}
R(\Theta)= R_{+}\left(1 +r \cos{\Theta}\right),
\label{Rtheta}
\end{equation}
where
\begin{equation}
 R_{+}=\frac{R(0)+ R(\pi)}{2};\hspace{0.5cm} r= \frac{R(0)-R(\pi)}{R(0)+R(\pi)}  >0,
\label{r}
\end{equation}
one finds that that the differential conductance $dI_0/dV < 0$  if
\begin{equation}
{\bar
D} <\frac{3r}{1+4r}
\label{criteria}
\end{equation}

%%%%%%%%%%%%%%%%%%%%%%%%%%%%%%%%%%%%%%%%%%%%%%%
%%%%%%%%%%%%%%%%%%%%%%%%%%%%%%%%%%%%%%%%%%%%%

In Section \ref{selfoscillations} we show that in the case of the S-shaped IVC $I_0(V)$ current and magnetization oscillations arise if the stack is incorporated in a current biased circuit in parallel with a capacitor (see Fig.\ref{circuit}).
 In this case the thermo-electonic control of the relative orientation of  layers 0 and 2 may be of two types depending on the ratio between the resistance of the stack $R(\Theta)$ and the resistance of  the rest of the  circuit $R_0$.

\begin{figure}
  %%%%%%%%%%%%%%%%%%%%%%%%%%%%%%%% \centerline{\psfig{figure=zlaser2.eps,width=8cm}}
 %\centerline{\includegraphics[width=0.65\columnwidth]{th-Ist.eps}}
  %\vspace{1cm}
\epsfig{file=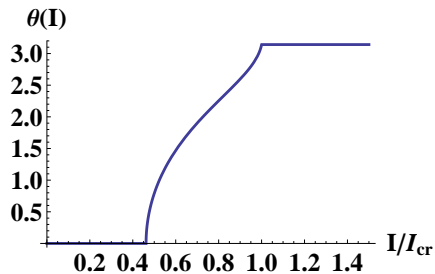, width=0.65\columnwidth} 
  \caption{The angle $\Theta$, which describes the tilt of the direction of
  the magnetization in layer 2 with respect to layer 0,
  as a function of the current in the current-biased regime. The curve was
  calculated for $R(\Theta)=R_+(1+r \cos\Theta)$, $R(\pi)/R(0)=0.3$, ${\bar
D}=0.36$; $I_{cr}=\sqrt{Q(T_c^{(1)})/R(0)}$.}
%The top and bottom graphs correspond to
%the voltage-bias and  the current-bias regimes, respectively.
 \label{theta-I}
 \end{figure}

As one can see from Fig.\ref{cvc}, if the bias current $I_{bias}$ through the stack is controllably  applied (that is $R_0 \gg R(\Theta)$),  the voltage drop across the stack is uniquely determined from the IVC $I_{bias}=I_0(V)$
and hence the relative orientation of the
magnetization of layers 0 and 2 (that is the tilt angle $\Theta$) can be changed smoothly from
parallel ($\Theta=0$) to anti-parallel $\Theta=\pi$ by varying the bias current through the
interval $I_{or} \leq  I_{\rm bias}\leq I_{cr}$. This corresponds to
moving along the $a-b$ branch of the IVC. The dependence of the
magnetization direction $\Theta$ on the current flowing through  the
stack is shown in Fig.~\ref{theta-I}.
%%%%%%%%%%%%%%%%%%%%%%%%%%%%%%%%%%%%%%%%%%%%%%%6
  %%%%%%%%%%%%%%%%%%%%%%%%%%%%%%%%%%%%%%%%%%%%%%%
  %%%%%%%%%%%%%%%%%%%%%%%%%%%%%%%%%%%%%%%%%%%%%%%

In the voltage-bias regime, on the other hand,  where the resistance of the stack is much larger than the resistance of the rest of the circuit,  $R(\Theta) \gg R_0$,  the voltage across the stack $V$ is kept at a given value  approximately equal to the bias voltage. Since the IVC is S-shaped, the stack
can now be in a bistable state: in the voltage range between points $a$ and $b$  there are three possible values of the current for one fixed value  of voltage (see Fig. \ref{cvc}). The states of the stack with the lowest and the highest currents are stable, while the state of the stack with  the middle value of the current is unstable. Therefore, a change of the voltage results in a
hysteresis loop: an increase of the
voltage along the $0-b$ branch of the IVC leaves the magnetization
directions in the stack parallel ($\Theta=0$) up to point $b$,
where the current  jumps to the upper branch $a-a'$, the jump
being accompanied by a fast switching of the stack magnetization
from the parallel ($\Theta=
0$) to the antiparallel orientation ($\Theta=\pi$). A decrease of the voltage along the $a'-a $ IVC branch keeps
the stack magnetization antiparallel up to point $a$, where the
current jumps to the lower $0-b$  branch of the IVC  and the magnetization of the stack returns to the parallel orientation  ($\Theta=0$).

In the next section we show that if the stack is connected in parallel with a capacitor, and the capacitance exceeds some critical value, the above time independent state becomes unstable and spontaneous temporal oscillations appear in the values of the current, voltage across the stack, temperature,
and direction of the magnetization.

\section{ Self-excited electrical, thermal and directional
magnetic oscillations. \label{selfoscillations}}
\begin{figure}[htbp]
%\centering \includegraphics[width=0.65\columnwidth]{Model.eps}
\epsfig{file=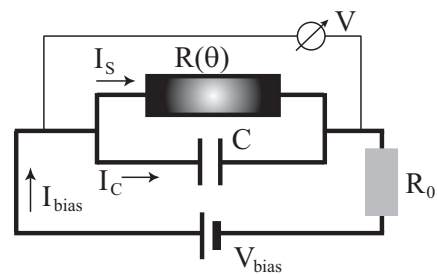, width=0.65\columnwidth} 
\caption{ Equivalent circuit for a Joule-heated magnetic stack of
the type shown in Fig.~\ref{noflip}.
  A  resistance $R[\Theta(t)]=I(t)/V(t)$, biased by a fixed DC current $ I_{\rm bias}$, is
  connected in parallel with a
  capacitor  $C$; $R(\Theta)$ and $R_0$ are the angle dependent resistance of the stack and the resistance of the rest of the circuit, respectively;  $I(\Theta)$ and $I_c$ are the currents flowing through the
  stack and the capacitor, respectively.} \label{circuit}
\end{figure}

We consider now a situation
where the magnetic
stack under investigation is incorporated into an electrical circuit in parallel with a capacitor of capacitance
$C$ and the circuit is biased by a DC current $I_{\rm bias}$, illustrated by the equivalent circuit of Fig.~\ref{circuit}.
The thermal and
electrical processes in this system are governed by the set of
equations
\begin{eqnarray}
C_V\frac{dT}{dt} &+ &Q(T) -R^{-1}(\Theta) V^2 =0;
\nonumber \\
C \frac{dV}{dt} &+&R^{-1}(\Theta) V= I_{\rm bias}\,, %\hspace{1.7cm}
 \label{evolution}
\end{eqnarray}
where $C_V$ is the heat capacity. The relaxation of the magnetic
moment to its thermodynamic equilibrium direction is assumed to be the fastest process in the problem,  which implies that  the magnetization direction corresponds to the equilibrium state of the stack at the given temperature $T(t)$. In other words,
the  tilt angle
$\Theta=\Theta(T(t))$ adiabatically follows the time-evolution of the temperature and hence its temperature dependence is given by Eq.~(\ref{theta2}).
%%%%%%%%%%%%%%%%%%%%%%%%%%%%%%%%%%%%%%%%%%%%%%%%%%%%%%%%%%%%%%7

%%%%%%%%%%%%%%%%%%%%%%%%%%%%%%%%%%%%%%%%%%%%%%%%%%%%%%%%%8
A time dependent variation of the temperature is accompanied by a
variation of the tilt angle $\Theta(T(t))$  and hence by
a change in the voltage  via the dependence of
the magneto-resistance on this angle, $R=R(\Theta)$.

The system of equations Eq.(\ref{evolution})  has one time-independent  solution ($\bar{T}(I_{bias}), \hspace{.1cm} \bar{V}(I_{bias}) $)
which is determined by the equations
\begin{equation}\label{SteadySolution}
R^{-1} \left[\Theta ( T)\right]V^2=Q( T),\hspace{0.5cm} R^{-1}\left[\Theta (T)\right]V=I_{bias}.
\end{equation}
 This solution is identical to the solution of Eqs.(\ref{theta2},\ref{heat}) that determines the S-shaped IVC shown in Fig.\ref{cvc} with a change $I\rightarrow I_{bias}$ and  $V\rightarrow  \bar{V}$.

In order to investigate the stability of this time-independent solution we write the temperature, current and angle as a sum of two terms,

\begin{eqnarray}
T= \bar{T}(I_{bias})+T_1(t); \nonumber \\
V=\bar{V}(I_{bias})+V_1(t); \nonumber \\
\Theta=\bar{\Theta}(I_{bias}) +\Theta_1(t),
\label{eqlinear}
\end{eqnarray}

where $T_1$, $V_1$ and $\Theta_1$ are small corrections. Inserting Eq.(\ref{eqlinear}) into Eq.(\ref{evolution}) and Eq.~(\ref{theta2}) one easily finds that the time-independent solution Eq.(\ref{SteadySolution})
is always stable at any value of the capacitance $C$ if the bias current $I_{bias}$ corresponds to a branch of the IVC with a positive differential resistance (branches a-a' and 0-b in Fig.\ref{cvc}).
If the bias current $I_{bias}$ corresponds to the branch with a negative differential resistance ($I_{or}<I_{bias}<I_{cr}$, see Fig.\ref{cvc}), the solution of the set of linearized equations is  $T_1=T_1^{(0)} \exp\{\gamma t\}$, $V_1=V_1^{(0)}\exp\{\gamma t\}$ and $\Theta_1=\Theta_1^{(0)}\exp\{\gamma t\}$ where $T_1^{(0)} $, $V_1^{(0)}$ and $\Theta_1^{(0)}$ are any initial values close to the time-independent state of the system, and
\begin{equation}
\gamma=\frac{1}{2\bar{R} C}\Biggl( \frac{C-C_{cr}}{C_{cr}}  \pm \sqrt{\left(\frac{C-C_{cr}}{C_{cr}}\right)^2-4\frac{\bar{R}}{|R_{d}|}\frac{C}{C_{cr}}}\, \Biggr),
 \label{gamma}
\end{equation}
where
\begin{equation}
C_{cr}=\frac{C_V}{|d(R Q)/dT|}\Bigl|_{T=T(V)},
 \label{criticalinductance}
\end{equation}
$R_{d}=d V/dI$ is the differential resistance, $\bar{R} = R(\bar{\Theta})$.

As one can see from Eq.(\ref{gamma}), the time-independent state Eq.(\ref{SteadySolution}) looses its stability if the capacitance exceeds the critical value $C_{cr}$, that is $C>C_{cr}$. In this case  a limit cycle appears in  plane ($V, T$)
(see, e.g., Ref. \cite{Andronov}),
associated with the arising self-excited, non-linear, periodic oscillations in temperature, $T=T(t)$, and voltage, $V=V(t)$. These are accompanied by oscillations of the  current through the stack, $I_s(t)=V(t)/R[\Theta(t)]$, and of the magnetization direction $\Theta(t) = \Theta(T(t))$. In the case when $(C-C_{cr})/C_{cr} \ll 1$, the system undergoes nearly harmonic oscillations around the steady state (see Eq.(\ref{eqlinear})) with frequency $\omega =\rm Im \left[ \gamma (C=C_{cr}) \right]$. Therefore, the temperature, $T$, the voltage drop across the stack, $V$, the current through the stack, $I_s(t)$, and the magnetization direction, $\Theta$,  perform a periodic motion with frequency
\begin{equation}
\omega = \frac{1}{C_{cr}\sqrt{\bar{R} |R_d|}}.
\label{frquency}
\end{equation}

With a further increase of the capacitance the size of the limit cycle grows, the amplitude of the oscillations increases and the oscillations become  anharmonic, with their period decreasing with increasing capacitance.

In order to investigate the time evolution of the voltage drop  and
the current through the stack  in more detail it is convenient to change variables $(V(t),T(t))$ to $(V(t), {\tilde I}(t))$ and introduce auxiliary
current  ${\tilde I}(t)$  and voltage $ V_0(t)$ related to
each other through Eqs.~(\ref{heat}) and (\ref{theta2}). Thus, we define
\begin{equation}
 {\tilde I}(t)= \sqrt{\frac{Q(T(t))}{R(T(t))} };\;\ \; V_0={\tilde
 I}(t)R(T(t))\,,
%\nonumber \\
 \label{VtildeJ}
\end{equation}
where $R(T)=R(\Theta(T))$. Comparing these expressions with
Eq.~(\ref{heat}) shows that at any moment $t$
Eq.~(\ref{VtildeJ}) gives the stationary IVC of the stack, $V_0= V_0({\tilde I})$, which is an inverse function of the current-voltage characteristic $I_0(V)$  defined by Eq.~(\ref{IVC0}) and shown in Fig.~(\ref{cvc}).

Differentiating ${\tilde I}(t)$ with respect to $t$  and using
Eqs.~(\ref{evolution}) and (\ref{VtildeJ}) one finds that the
dynamic evolution of the system is governed by
\begin{eqnarray}
\tau_0 \frac{d {\tilde I}}{d t}-\frac{V^2 -V_0^2({\tilde
I})}{2V_0({\tilde I})}=0,
\nonumber \\
C \frac{d  V}{d t}+ \frac{{\tilde I}}{V_0({\tilde
I})}V=I_{\rm bias}
 \label{evolutionJV}
\end{eqnarray}
where $$\tau_0=  \frac{C_V }{d(QR^{-1})/dT}\Bigr|_{T=T({\tilde V})}.$$ Eq.~(\ref{VtildeJ}) indicates that at  any
moment $t$ the current through  the stack $I_s(t)=V(t)/R(T(t))$ is
coupled with the auxiliary voltage ${\tilde V}(t)$ by the following relation:
$$I_s= \frac{V}{V_0({\tilde V})}{\tilde I}.$$

%%%%%%%%%%%%%%%%%%%%%%%%%%%%%%%%%%%%%%%%%%%%%%%%%%%%%%%%%9
 \begin{figure}
%  \centerline{\includegraphics[width=0.65\columnwidth]{I-Vdy.eps}}
  %\vspace{1cm}
\epsfig{file=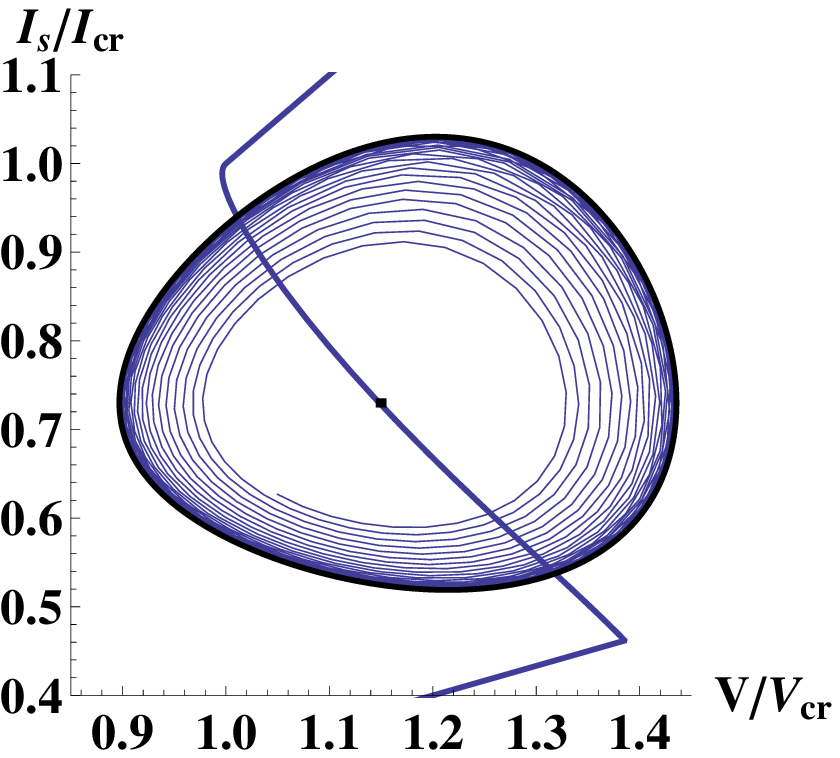, width=0.65\columnwidth}
  \caption{Spontaneous oscillations of the current through the stack, $I(t)$,
 and the voltage drop across it, $V(t)$, calculated for
  $R(\pi)/R(0)=0.3$, ${\bar
D}=0.36$ and
$(C -C_{\rm cr})/C_{\rm cr}=0.062$; $I_{cr}=\sqrt{Q(T_c^{(1)})/R(\pi)}$ and
$V_{cr}=R(\pi)I_{cr}$.  $I(t)$ and $V(t)$
develop from the initial state toward the limiting cycle shown by the thick solid line, along which they execute a periodic motion. The stationary IVC of the stack is shown by the thin solid line.}
   \label{cvclimmiddle}
  \end{figure}
%%%%%%%%%%%%%%%%%%%%%%%%%%%%%%%%%%%%%%%%%%%%%%%

The coupled equations (\ref{evolutionJV}) have only one  time-independent
solution  $V=V_0(I_{bias})$ where $V_0(I)$ is the voltage-current characteristic (its inverse function $I_0(V)$ is shown in Fig.\ref{cvc}). However, in
the interval $I_{or} \leq I_{\rm bias} \leq I_{cr}$ this solution is
unstable with respect to small perturbations if $C > C_{\rm cr}$. As a result periodic oscillations of $I(t)$ and $ {\tilde V}(t)$ appear spontaneously, with
$I(t)$, $ {\tilde V}(t)$ moving along a limit cycle. The limit cycle in the
$I$-$V$ plane is shown in Fig.~\ref{cvclimmiddle}.
 % and \ref{cvclimbig}.

\begin{figure}
%  \centerline{\includegraphics[width=0.65\columnwidth]{th_t.eps}}
  %\vspace{1cm}
\epsfig{file=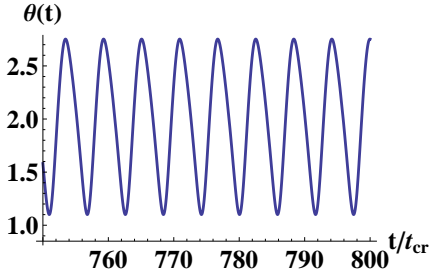, width=0.65\columnwidth}
  \caption{Spontaneous oscillations of the magnetization direction
  angle, $\Theta(t)$,
   calculated for
  $R(\pi)/R(0)=0.3$, ${\bar
D}=0.36$ and
$(C -C_{\rm cr})/C_{\rm cr}=0.062$.}
   \label{limthetasmall}
  \end{figure}

 Eq.(\ref{theta2}) and Eq.(\ref{VtildeJ}) show that the magnetization direction, $\Theta(t)= \Theta(T(t))$, and  the stack temperature
 $T=T(t)$
 %and  voltage drop over the stack $V(t)=J(t) R(\Theta(t))$
 follow these electrical oscillations adiabatically according to the
 relation $$Q[T(t)]= {\tilde I}(t)V_0[{\tilde I}(t)]. $$  Temporal oscillations of $\Theta(t)$ are shown in  Fig.~\ref{limthetasmall}.

\begin{figure}
%  \centerline{\includegraphics[width=0.65\columnwidth]{I-VdyhC.eps}}
  %{limIVcyclbig1_new.eps}
  %\vspace{1cm}
\epsfig{file=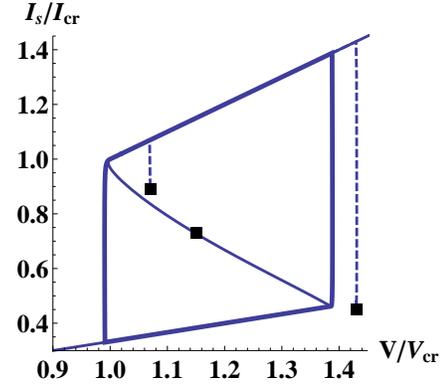, width=0.65\columnwidth}
  \caption{Spontaneous oscillations of current,
$I(t)$,
  and voltage, $V(t)$,  calculated for  $R(\pi)/R(0)=0.3$, ${\bar
D}=0.36$  and
$C/C_{\rm cr}=520.91$. $I(t)$ and $V(t)$
develop toward the limit cycle (thick solid line) from the initial state which can be either inside or outside it, and execute a periodic motion along the limit cycle. The stationary IVC of the stack is shown as a thin solid line.}
   \label{limIVbig}
  \end{figure}

\begin{figure}
%  \centerline{\includegraphics[width=0.65\columnwidth]{th_t_hc.eps}}
  %{limIVcyclbig1_new.eps}
  %\vspace{1cm}
\epsfig{file=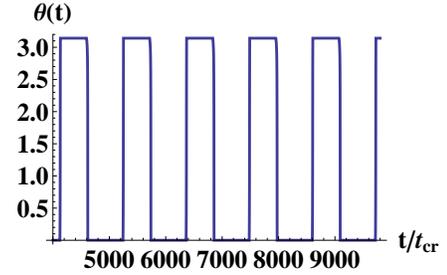, width=0.65\columnwidth}
  \caption{Spontaneous oscillations of the magnetization direction
  angle, $\Theta(t)$,
   calculated for
  $R(\pi)/R(0)=0.3$, ${\bar
D}=0.36$  and
$C /C_{\rm cr}=520.91$.}
   \label{limthetabig}
  \end{figure}

 According to Eq.(\ref{evolutionJV}) the ratio between the characteristic evolution times of the voltage $V(t)$ and current ${\tilde I}(t)$ increases with an increase of the capacitance, resulting in a qualitative change of the character of oscillations  in the limit $C \gg C_{\rm cr}$.
 In this case the current and the voltage slowly
move along branches $a - a'$  and $0 - b$
 of the IVC (see Fig.\ref{cvc})) at the rate ${\dot V}/V \approx
1/R_+ C$ (here ${\dot V} =d V/d t$),
%${\dot J}/J
%\approx 1/t_J$ (here $t_J ={\cal L}/R_+$ is the
%  current evolution time)
  quickly switching between these branches
at points $a$  and $b$  with at the rate of $\sim
R_+/\tau_0$ (see Fig.~\ref{limIVbig}). Therefore, the stack in this limit
periodically switches between the parallel and antiparallel magnetic
states, as shown in Fig.~\ref{limthetabig}.
%%%%%%%%%%%%%%%%%%%%%%%%%%%%%%%%%%%%%%%%%%%%%%%%%%%%%%%%%12
  %%%%%%%%%%%%%%%%%%%%%%%%%%%%%%%%%%%%%%%%%%%%%%%

%%%%%%%%%%%%%%%%%%%%%%%%%%%%%%%%%%%%%%%%%%%%%%%%%%%%%%%%%11
  %%%%%%%%%%%%%%%%%%%%%%%%%%%%%%%%%%%%%%%%%%%%%%%
%%%%%%%%%%%%%%%%%%%%%%%%%%%%%%%%%%%%%%%%%%%%%%%%%%%%%%%%%11
  %%%%%%%%%%%%%%%%%%%%%%%%%%%%%%%%%%%%%%%%%%%%%%%
Using equations Eq.(\ref{SteadySolution})  one can estimate the order of magnitude  of the critical capacitance $C_{cr}$,  Eq.(\ref{criticalinductance}),   and the oscillation frequency $\omega$, Eq.(\ref{frquency}),  as $C_{cr}\approx   T c_v d/(\rho j)^2$ and $\omega \approx \rho j^2/T c_v$, where $c_v$ is the heat capacity per unit volume, $\rho$ is the  resistivity, and $d$ is the characteristic size of the stack. For point contact devices with   typical values of $d \sim 10^{-6} \div 10^{-5} \rm cm$, $c_v \sim 1\; \mathrm{J/cm}^3$K, $\rho\sim 10^{-5} $ $\Omega$cm,  $j\sim 10^8 \;\mathrm{ A/cm^2}$ and
assuming that cooling of the device can provide the sample temperature $T\approx T_c^{(1)}\sim 10^2 \rm K$ one finds for the characteristic values of the critical capacitance and the oscillation frequency
$C_{\rm cr}\approx 10^{-11} \div 10^{-10}\rm F$ and $\omega \approx 1 {\rm GHz}$, respectively.

\begin{figure}
%  \centerline{\includegraphics[width=0.65\columnwidth]{STE-C_exp_fig.eps}}
\epsfig{file=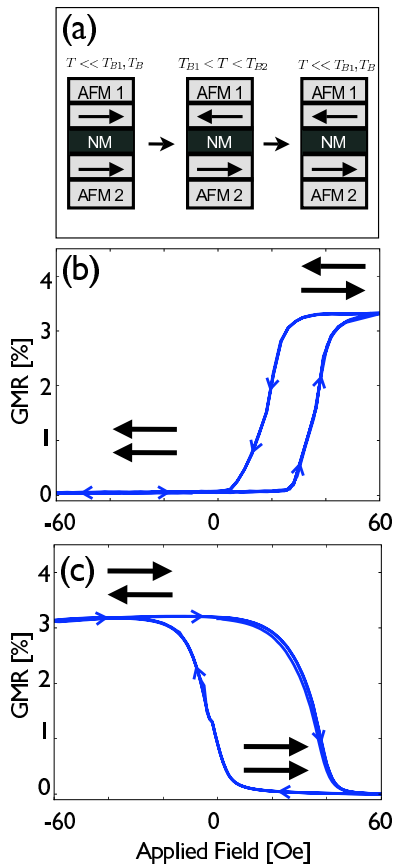, width=0.65\columnwidth}
  \caption{(a) Schematic of the studied material system: two ferromagnets, FM 1 and FM 2, each exchange pinned by an antiferromagnet, AFM 1 and AFM 2, are separated by a non-magnetic alloy, NM. The middle and right panels illustrate two methods to alter the exchange pinning such that the magnetization of the two ferromagnets become antiparallel. $T_{B}$ is the antiferromagnetic blocking temperature and $H_{C}$ is the coercive field of the ferromagnets. (b) and (c) are minor magnetoresistance loops before and after the field-heat treatment, respectively. Arrows indicate the magnetization direction in FM 1 and FM 2.}
   \label{experiment}
  \end{figure}

We will now show how the material system discussed above, with inverse GMR, can be fabricated using differential exchange-biasing. A schematic of the layer structure is shown in Fig.~\ref{experiment} (a). The two ferromagnetic layers of the spin-valve (FM 1 and FM 2), each exchange pinned by an antiferromagnet (AFM 1 and AFM 2), are separated by a non-magnetic metal spacer (NM). Normally, the exchange-biasing procedure performed at a high field results in the two ferromagnets being pinned with their magnetization parallel to each other. In order to demonstrate inverse magnetoresistance the exchange pinning must be altered such that the magnetization of the two ferromagnets become antiparallel. This can be done in two ways.

%%%%%%%%%%%%%%%%%%%%%%%%%%%%%%%%%%%%%%%%%%%%%%%%%%%%%%%%%11
  %%%%%%%%%%%%%%%%%%%%%%%%%%%%%%%%%%%%%%%%%%%%%%%

First; the blocking temperature of the two antiferromagnetic layers can be designed in fabrication such that one is lower than the other, $T_{B1} < T_{B2}$  or $T_{B2} < T_{B1}$ \cite{ref:ThermalMRAM}. By heating the sample to a temperature above the lower blocking temperature $T_{B1}$ ($T_{B2}$) but below the higher blocking temperature $T_{B2}$ ($T_{B1}$), the magnetization of FM 1 (FM 2) can be rotated by applying an external field. With FM 1 and FM 2 now antiparallel the sample is cooled down to room temperature and the two ferromagnetic layers are pinned in opposition.

Second; by using ferromagnets with different coercivity, $H_{C1} \neq H_{C2}$, and heating the sample to a temperature above both $T_{B1}$ and $T_{B2}$. At this temperature the ferromagnet with lower coercivity, FM 1 (FM 2), can be rotated in an external field if $H_{C1} < H_{A} < H_{C2}$ ($H_{C2} <  H_{A} < H_{C1}$), where $H_{A}$ is the applied field. With FM 1 and FM 2 now antiparallel the sample is cooled down to room temperature. A schematic of these two methods are shown in Fig.~\ref{experiment} (a). Below, we demonstrate the second method for obtaining an inverse GMR spin-valve using a practical material stack.

A series of films were deposited on thermally oxidized Si substrates using magnetron sputtering at a base pressure better than $5 \cdot 10^{-8}$ Torr. 15 nanometer thick IrMn is used for AFM 1 and AFM 2. FM 1 is a thin bilayer of CoFe/NiFe and FM 2 is made up of CoFe/NiCu/CoFe/NiFe/CoFe, such that the two layers have different coercivity ($H_{C1} > H_{C2}$). To induce exchange coupling at the AFM/FM interfaces the deposition was carried out in a magnetic field higher than 350 Oe. The complete structure of the sample was NiFe 3/ IrMn 15/ CoFe 2/ NiCu 30/ CoFe 2/ NiFe 6/ CoFe 4/ Cu 7/ CoFe 5/ NiFe 3/ IrMn 15/ Ta 5, with the thicknesses in nanometers.

Using a vibrating sample magnetometer (VSM) equipped with an oven, the magnetization and switching fields of FM 1 and FM 2 could be monitored while the temperature was increased. At 180$^{\circ}$C the temperature was higher than both $T_{B1}$ and $T_{B2}$ which made it possible to align FM 1 and FM 2 antiparallel (since $H_{C1} > H_{C2}$). The external magnetic field was then removed and the sample was cooled down to room temperature. The minor magnetoresistance loop (current in plane magnetoresistance versus applied field when only the magnetization of FM 2 is switched) before and after the above field-heat treatment is shown in Fig.~\ref{experiment} (b). It can be seen that before the heat treatment the resistance increases when a positive field of 30 Oe is applied. The same measurement for the same sample but after the heat treatment is shown in Fig.~\ref{experiment} (c). The magnetoresistance has now become inverse which means that the resistance decreases when a positive field is applied. The coercivity of FM 2 chosen for this demonstration of differential exchange pinning has increased somewhat as a result of the field-heat treatment. This issue should yield to optimization of the material stack and working with smaller blocking temperatures.

As regards to the size of the integrated oscillator, let us assume the typical resistance of the STE-valve to be 50 Ohm and the operating frequency $\omega  =1$ GHz, and estimate the required inductance ${\cal L}$ and capacitance $C$ of the circuit. Thus, $\omega {\cal L}=50$ Ohm yields ${\cal L} \sim 10^{-7}$ H while $1/(\omega C)=50$ Ohm yields $C \sim 10^{-11}$ F. The typical inductance of an on-chip inductor is 1 nH per mm of the wire length, somewhat enhanced if the wire is turned in to a spiral. Therefore the required chip area would be millimeters squared in size. A square parallel plate capacitor, estimated using the standard formula, for a typical dielectric material of a few nm thick insulator would measure 10 micrometers on the side. Obviously, a 10,000-fold reduction in the typical footprint should be achievable in principle by employing a device based on an S-shaped STE-valve. In systems where tunability and high power are desirable, the N-shaped inductor based design may be preferable.

\section{Conclusion}
We have shown that Joule heating of an exchange-spring nanopillar with an inverse magneto-resistance results in a mutual coupling of the current and  orientation of the switching layer. This allows to control the magnetic alignment in the stack in the full range of angles from parallel to antiparallel by varying the current through the structure. We have determined the range of parameters in which the current-voltage characteristic of the exchange-spring nanopillar with inverse magneto-resistance is S-shaped and evaluated a spin-thermionic oscillator whoose frequency can be varied by changing the capacitance in the circuit from essentially dc to the GHz range. A suitable material stack and method to produce the required inverse-GMR spin-valve is demonstrated.

\section{Acknowledgement.}
 Financial support from the European Commission (Grant No. FP7-ICT-2007-C; proj no 225955 STELE)
%, Swedish VR and SSF, and the Korean WCU programme funded by MEST through KOSEF (R31-2008-000-10057-0)
is gratefully acknowledged.


\begin{thebibliography}{99}
\bibitem{Giovanni} The orientational phase transition in such a system induced by an external magnetic field was considered in paper \cite{Asti}.
\bibitem{inverse} Condition $dR/d\Theta < 0$, which is necessary to obtain an S-shaped IVC, can also be obtained in the case where there is a potential barrier between layers 2 and 3.
\end{thebibliography}
\end{document}